# Vortex dynamics at the transition to the normal state in $YBa_2Cu_3O_{7-\delta}$ films

P. Bernstein [a)], J.F. Hamet [a)], M.T. González [b,*)] and M. Ruibal Acuña [b)]

[a)] CRISMAT-ENSICAEN (UMR-CNRS 6508) F14050 Caen cedex 4, France

[b)] LBTS, Departamento de Física da Materia Condensada Universidade de Santiago de Compostela E15782, Spain

**Abstract** : We propose a description of the vortex dynamics in $YBa_2Cu_3O_{7-\delta}$ films from the critical to the normal states. This description supposes that the vortex motion is thermally activated along the twin boundaries of the films. The discontinuity observed in the current-voltage curves at the transition to the normal state is explained by the sudden increase in the dissipated power rate due to vortex depinning. However, near the critical temperature, this phenomenon does not occur because the vortex activation energy is near zero. We also show how the current at the transition to the normal state can be computed from the current-voltage curves measured at low currents. The predictions of this description are compared to the data published by González *et al.* [Phys.Rev.B**68**,054514 (2003)].

Corresponding author : Pierre Bernstein

CRISMAT-ENSICAEN, Boulevard du Maréchal Juin, F14050 FRANCE

tel : 33 2 31452685

fax : 33 2 31951600

e-mail : pierre.bernstein@ensicaen.fr

# I - Introduction

Due to its importance in thin film fault current limiters, the process governing the current induced transition to the normal state of $YBa_2Cu_3O_{7-\delta}$ (YBCO) films has been widely discussed in the literature [1,2,3,4]. For many authors this transition must be mostly ascribed to the film vortex dynamics. In the Larkin-Ovchinikov (LO) model, the electric field generated by the vortex motion shifts upwards the energy distribution of the quasi-particles with respect to the equilibrium distribution [5,6,7]. As a consequence, the quasiparticles leave the vortex core, the viscous damping coefficient of the vortices is reduced and the vortex velocity increases, until an instability point is reached. However, according to Xiao and Ziemann [8], the LO model alone does not account for all the features of the transition in YBCO films and they claim that an additional depinning mechanism is involved. In another paper [9], they suggest that the behavior of the films before and at the transition is the effect of the self organized criticality (SOC) of the vortex lines. According to this proposition, the discontinuity at the transition is due to the jump of a vortex line that triggers a chain reaction in the vortex lines or an avalanche. This model is supported by computer simulations [10], magneto-optical observations [11] and theoretical calculations [12]. Other authors suggest that the transition is due to the generation of phase slip centers (PSCs). In principle this phenomenon occurs in 1D systems only, however several groups have claimed that they could observe PSCs in films [13,14,15]. At the location of a PSC the modulus of the order parameter vanishes periodically with the Josephson frequency while the phase changes by $2\pi$. Then, the area of the PSC turns resistive and the voltage experiences a discontinuity. The PSCs first remain localized but they transfer heat to the sample and, for a large enough current, they expand and drive the whole sample in the normal state. As a result, the current-voltage curves (CVCs) show several small voltage jumps, each corresponding to the ignition of a determined PSC, before the transition to the normal state occurs. A puzzling aspect of the transition to the normal state is that it is continuous in the near vicinity of $T_c$, the critical temperature, while it is discontinuous at lower temperatures. Then, Xiao *et al.* in Ref.[16] claim that two different processes occur in YBCO films. At low temperature, discontinuities in the CVCs would be due to the generation and the expansion of hot spots while the transition in the vicinity of $T_c$ could be attributed to a LO process. According to the models described above, the transition is not due to thermal instability, as proposed for example by Peterson *et al.*[17].

Gonzàlez *et al.* [18] have measured the CVCs with no applied field of c-axis oriented thin films deposited on $SrTiO_3$ substrates and those of melt-textured samples in the whole range of current from the non dissipative to the normal states. They have shown that the CVCs measured on both types of samples can be fitted with the same expression they named the critical power law (CPL). In addition, if $J_c$ and $J^*$ are the critical current density and the current density at the transition to the normal state respectively, they found that the values of $J^*/J_c$ computed for the melt textured samples show a dependence on temperature very similar to those of thin films. These arguments support the suggestion that intrinsic vortex dynamics plays the main role in the transition to the normal state. However, from other works carried out by the same group [19, 20], the authors show that in films the discontinuity at the transition to the normal state can be reproduced satisfactorily considering the effect of thermal heating only.

Starting from a different point of view, Bernstein and Hamet have proposed a description of the vortex dynamics accounting for the transport properties of YBCO films in the vicinity of the critical state [21]. According to this description, most of the dissipation in the resistive state is due to the motion of vortices along the twin boundaries (TBs) of the films. Due to the presence of numerous defects, the TBs split off into rows of weak links that behave as Josephson junctions and carry the same current in the critical state.

In this paper, we develop that model and we show that it can account for the results of González *et al.*[18], especially the differences observed at high current in the CVCs measured at different temperatures and the values of J*. The paper is organized as follows. In section II we remind of the main results of the model in Ref.[21] and we present developments that account for the behavior of the CVCs at the transition. We propose a mechanism based on vortex interaction for the ignition of the transition to the normal state. We show how it is possible to infer the value of J* at a given temperature from the CVC measured in the vicinity of $J_c$ at the same temperature. In section III we compare the experimental results obtained by González *et al.* to the model predictions. Section IV is devoted to a discussion.

**II – Vortex dynamics in the critical state and at the transition to the normal state**

In this section, we present and develop the main aspects of the vortex dynamics according to the model in Ref.[21], we propose a description of the change in the vortex regime occurring at J* and we detail how this last quantity can be computed.

**II – A  Vortex dynamics in the critical state**

In the critical state the vortex motion in YBCO films is thermally activated. The CVCs can be described with the general Kim-Anderson expression

$$E = E_o \exp\left(-\frac{U_o}{k_B T}\right) \sinh\left(\frac{W}{k_B T}\right) \qquad (1)$$

where E is the electric field measured at the sample terminals, $U_o$ the zero current vortex activation energy or pinning energy, W the work carried out by the current and $E_o$ a constant.

In the simplest form of the Kim-Anderson model the quantity $U(J)= U_o-W$ is a linear function. However, many authors have suggested that U(J) takes other forms. From collective pinning calculations Feigel'man *et al.*[22] have claimed that $U(J) = U_o \left(\frac{J_c}{J}\right)^\mu$ where $\mu$ depends on the size of the moving vortex bundles and is equal to 1/7 for isolated vortices. Zeldov *et al.*[23] have suggested that, due to the shape of the pinning wells, U(J) has a logarithmic form that yields the power law $E(J) = E_o \left(\frac{J}{J_c}\right)^n$ with $n = \frac{U(T,B)}{k_B T}$, where B is the applied magnetic field. As a general rule, extensions of the Kim-Anderson model are based on some hypotheses on the vortex dynamics and the mechanisms responsible for pinning. The role of extended planar defects in the physics of YBCO, i.e. low angle grain boundaries (LGBs) and twin boundaries, has been discussed for a long time [24,25]. Many authors have pointed out that since the width of these planar defects in YBCO films is in the range of the superconducting coherence length in the a-b planes, some type of Josephson behavior could be expected. Gurevitch and Colley [26] consider the behavior of vortices in presence of a network of planar defects parallel to the flux lines and claim that they are pinned by the non uniformity of the maximum Josephson current flowing across these planes. Mezzetti *et al.* consider that the boundary planes behave as long Josephson junctions whose coupling energy is modulated by defects [27].

Our model considers the properties of YBCO films epitaxially grown on $SrTiO_3$ substrates. According to electron microscopy observations, the TBs can stretch over several micrometers in this type of samples [28]. Magnetic susceptibility measurements suggest that the TBs planes act as grooves channeling the vortices and that vortex pinning occurs at the TBs intersections [29]. The vortex channeling effect of the TBs was previously pointed out by different types of measurements [30,31]. Then, we assume in this paper that the defects that are important for the transport properties of the films are the twin boundaries.

**1- Model description**

While structural investigations by common TEM (Transmission Electron Microscopy) show twin boundaries as regular extended defects associated to the presence of an important strain field [28, 32], HREM (High Resolution Electron Microscopy) observations reveal the non uniformity of the twin boundaries at the scale of a few interatomic distances [33,34]. These defects correspond to tiny atomic displacements, variation in atoms coordination or local vacancies. In a stoichiometric compound ($YBa_2Cu_3O_7$), the existence of TBs causes a variation of the copper coordination. The copper environment is no more a square but an octahedron or a tetrahedron. In non-stochiometric compounds ($YBa_2Cu_3O_{7-\delta}$) oxygen vacancies along the TBs avoid the high energy state of neighbouring oxygen atoms and preserve the coherency at the TBs locations. The average size of this disorder is three cells for the most coherent TBs. In addition, the existence of TBs is associated to a lattice rotation which is not 90° but around 89°1, which implies a lattice distortion and strain fields at the TBs locations that change locally the TBs width. Since the TBs width is in the range of the coherence length, it is reasonable to suppose that the modulation of this width by disordered ares has a strong effect on the TBs Josephson behavior.

We consider that when a current flows in a twinned YBCO film with no applied magnetic field, the TBs channel the vortices and the antivortices nucleated along each edge of the sample. As discussed above, the TBs width is in the range of $\xi_{ab}(T)$, the superconducting coherence lentgh in the a-b planes of YBCO, but they are highly disordered at the scale of a few interatomic distances. Then, it is reasonable to suppose that the separation between the superconducting banks of the TBs can be locally large enough with respect to $\xi_{ab}(T)$ to cause a disruption of the tunneling current. This results in the splitting of the TBs into rows of Josephson junctions or more probably of weak links (see Fig.1). Since $\xi_{ab}(T)$ is an increasing function of the temperature, more disordered areas acting as defects are effective for disrupting the Josephson current as the temperature decreases (see Fig.1). As a result, $\overline{\delta}$, the mean length of the weak links along the TBs is expected to be an increasing function of the temperature. When a bias current flows in a sample, the current and the flux first enter its peripheral part while the central area remains flux and current free for $J<J_c$ [35]. Magneto-optical observations have shown that the TBs are the first areas of the samples penetrated by the magnetic flux [36]. As a consequence, we assume that the TBs accommodate vortices whose screening currents flow across their weak links in the region penetrated by the flux. The area of the current and flux free central region decreases as the current increases. When this region disappears, vortices and antivortices are set in motion and annihilate each other in the middle of the sample along an annihilation line. Then, the film is in the critical state ($J=J_c$) and since a voltage can be measured at the film terminals, it enters the mixed state for $J>J_c$. In the critical state, all the weak links along the TBs are expected to carry current except those at the vortex cores. The strong disorder due to the defects in the weak links and considerations on the vortices and TBs energies suggest that in the critical state the weak link energy is equal to $k_BT$ (see Annex A). Then, all the weak links, whatever their length, carry the same net current that is equal to

$$I_J = \frac{2\pi k_B T}{\phi_o} \quad (2)$$

where $\phi_o$ is the flux quantum. An experimental verification of this prediction is reported in section III-D-1.

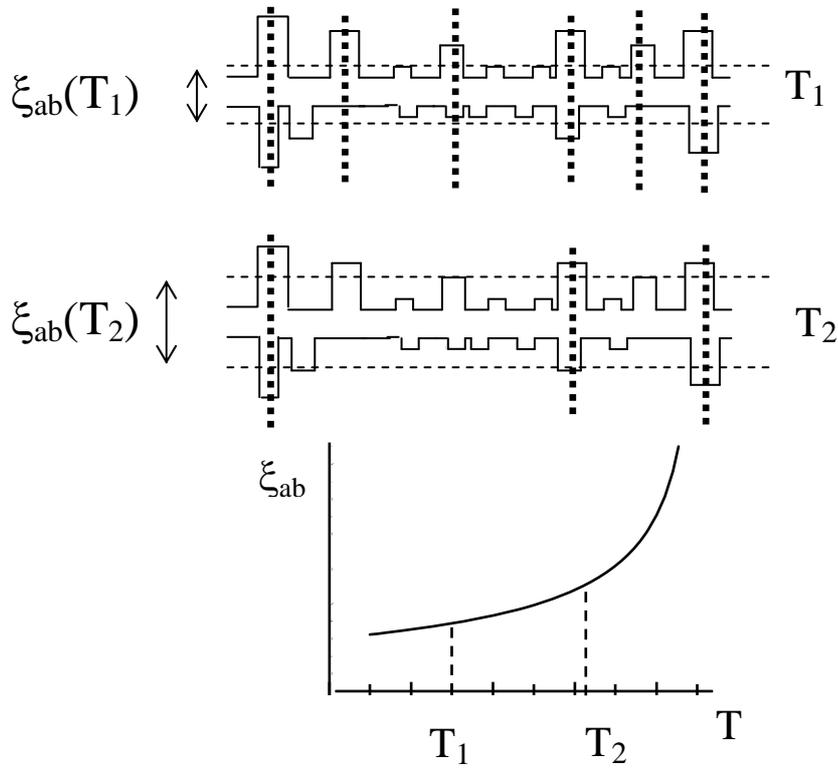

Fig.1 : Schematic representation of a twin boundary section at two different temperatures ($T_1 < T_2$) and sketch of the dependence on temperature of the coherence length, $\xi_{ab}$. Disordered areas lying along the twin boundaries bound weak links carrying a superconducting current. The dashed lines represent the limits of these weak links. The current is disrupted only if the effective separation between the superconducting banks is much larger than the coherence length. Since at temperature $T_1$ the coherence length is shorter than at temperature $T_2$, more defects are large enough to disrupt the superconducting current at $T_1$ than at $T_2$. Consequently, the weaks links are shorter at $T_1$ than at $T_2$.

If the sample is patterned as a strip, the mean length of the TBs paths connecting both strip edges is in the range of w, the strip width and $I_c$, the critical current of the strip can be written as

$$I_c = I_J \overline{N} \approx I_J \frac{w}{\delta} \qquad (3).$$

In Eq.(3), $\overline{N}$ is the mean number of weak links along a TBs path. Another consequence of Eq.(2) is that the amplitude of the vortex screening current flowing across each weak link is also at least equal to $I_J$, since a current amplitude smaller than $I_J$ is in the range of the current thermal fluctuations [37]. Dissipation occurs through thermally activated vortex motion along the TBs. A possible process for the vortex motion was described in Fig.8 of Ref.[21]. In this contribution, we propose a more probable scenario based on vortex pinning at the TBs intersections, in agreement with the observations of Berger *et al.*[29]. Fig.2a represents schematically two perpendicular vortex rows in the vicinity of a TB intersection. We assume that in the critical state there is no interaction between the vortices located on the same TB. However, the vortices of two intersecting TBs interact as they approach the intersection (see Fig.2b). The vortex in the vertical row can move forward over distance $\delta$ only if screening current lines enter both its core and the core of the first vortex in the horizontal row. Since the intensity of the vortex screening current flowing across a weak link is equal to $I_J$, the corresponding energy barrier takes the form

$$U_o = 2I_J \nu \phi_o = 4\pi \nu k_B T \qquad (4)$$

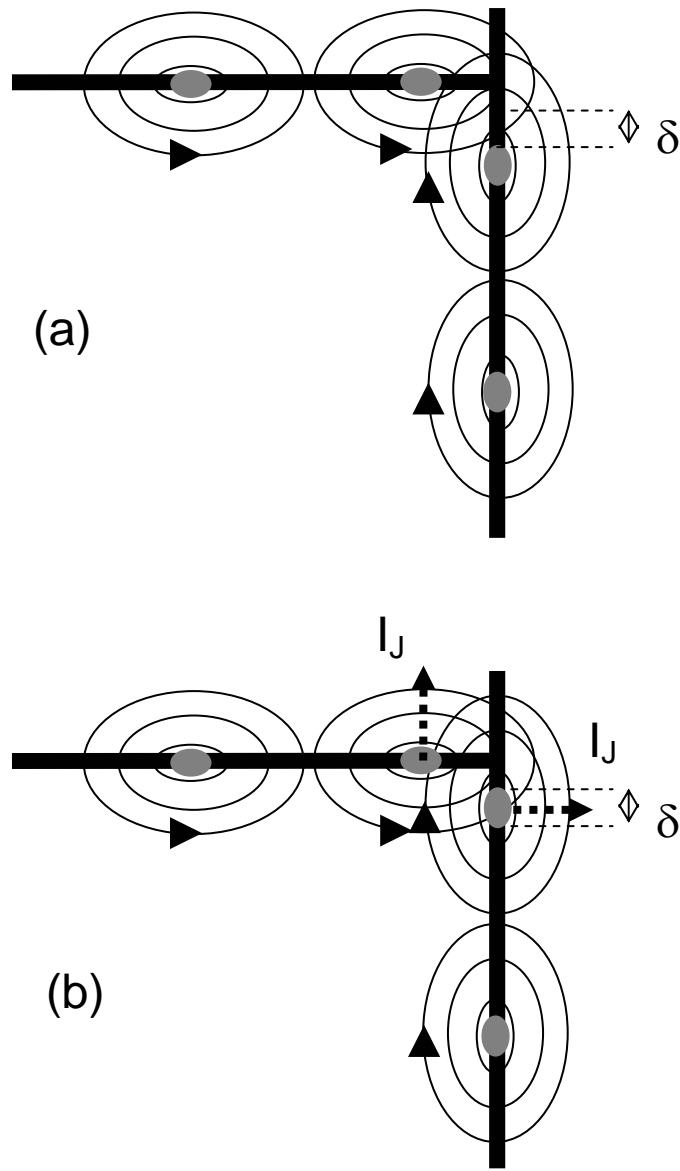

Fig.2 : a) Schematic representation of vortices located along two intersecting twin boundaries. The thick lines, the grey areas and the ellipses represent the twin boundaries, the vortex cores and the screening current lines, respectively; b) the first vortex in the vertical row can move forward over distance δ only if screening current $I_J$ enters both its core and the core of the first vortex in the horizontal row.

where $\nu\phi_o$ is the flux carried by each vortex. The suggestion that the vortex pinning energy takes the form of Eq.(4) is supported by measurements that have been carried out in a large range of temperature on twinned YBCO films, single crystals and powder grains by mutual inductance, magnetic relaxation and transport measurements [21, 38, 39, 40, 41, 42, 43]. Below an upper temperature, $T_{up}$, the quantity $\frac{U_o}{4\pi k_B T}$ computed from these measurements takes integer values, as expected from Eq.(4) [21]. The fact that, in some cases, $\nu$ is found different from unity is puzzling. In the case of films including columnar defects Buzdin has shown that a vortex can carry multiple flux quanta if some conditions are fulfilled [44]. He found that the creation of vortices with more than one flux quantum is energetically favorable if the condition $R^3 > \xi_{ab} a^2$ is fulfilled. Here a is the distance between two vortex cores and R is the defects radius. Multiquanta vortices have been observed in mesoscopic samples and in regular samples including an array either of holes or of magnetic particles, when this array is commensurate with the vortex lattice [45, 46]. In YBCO films the width of the twinned domains is in the 30nm range [29] but twin boundaries do not build a regular array. Then, commensurability effects are not expected. However, scanning SQUID and Hall microscopy of the flux generated by vortices trapped in YBCO films have yielded results suggesting that the magnetic flux generated by some vortices can be larger than $\phi_o$ [47, 48]. Although these observations can also be explained by the existence of single quantum vortex lines whose separation is abnormally short, they are not inconsistent with the suggestion that vortices can carry more than one flux quantum. To conclude, the solution to this problem requires probably a fine description of the physics of moving vortices constrained by a bias current to run along a TB including TBs intersections whose separation can be as short as a few coherence lengths.

**2-Domain of validity**

The model is not relevant above $T_{up}$ and at very low temperature. Above $T_{up}$, $U_o$ goes to zero as T goes to $T_c$ (see Fig.3 for an example). Considerations on the vortex dynamics above $T_{up}$ are proposed in section IV. At very low temperatures, $\xi_{ab}(T)$ and, as a result, $\overline{\delta}$ are almost constants. Then, the weak links carry a current larger than $I_J$ since the number of superconducting pairs increases as the temperature decreases, while the number of weak links along the TBs remains approximately constant. If an external field or a strong transport current is applied, we expect that the model is valid as far as the vortex dynamics is dominated by the motion of vortices along the TBs. This implies that the vortex interactions with the vortices either located on the same TB or nucleated in the bulk of the film can be neglected. As a result, the model fails if a large enough magnetic field is applied to the sample. Some authors have suggested that the planar defects in the films are not twin boundaries but low angle grain boundaries [49, 50]. Since we expect that the LGBs are at least as disordered as the TBs and show a Josephson behavior, they can probably be described with the same model as the TBs if the LGBs intersections look like Ts. In case the LGBs intersections are crosses, a vortex located along a LGB near an intersection is expected to interact with two vortices of the intersecting LGB and the energy barrier to take the form $U_o = 8\pi\nu k_B T$.

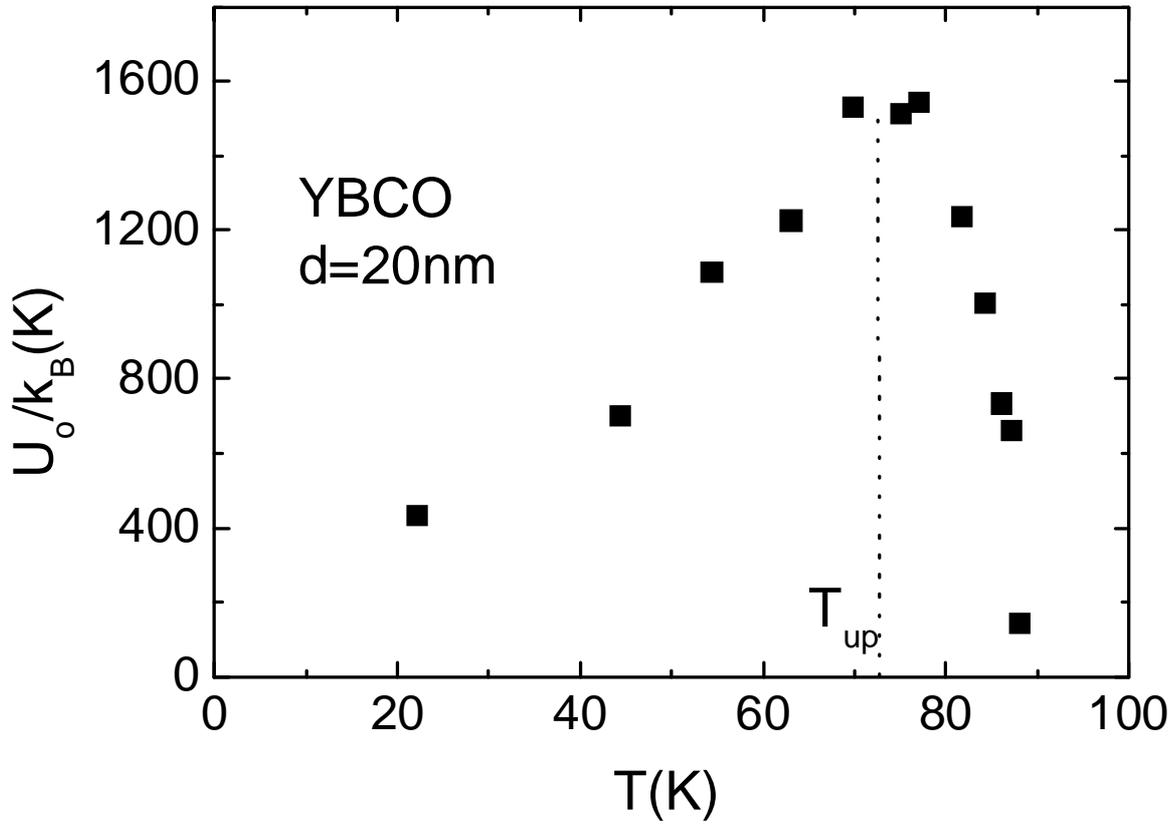

Fig.3 : Pinning energy of a 20 nm thick YBCO film deposited on a $SrTiO_3$ substrate as a function of the temperature. The pinning energy was computed from the current-voltage curves according to the method described in Ref.[21]. The pinning energy is an increasing function of the temperature below $T_{up}$ and goes to zero as T goes to $T_c$.

## II – B  Transition to the normal state

When transport current I flows in a strip, the dissipated power due to the vortex motion takes the form

$$P = IV = I n_m L \bar{v} \phi_o \quad (5)$$

where L is the length of the strip, V the voltage measured at its terminals, $n_m$ the mean surface density of the moving vortices and $\bar{v}$ their mean velocity. Since the vortex motion is thermally activated, it is reasonable to expect that if the current density in the strip is smaller than J*, $\bar{v}$ takes the form

$$\bar{v} = f(J) e^{-\frac{U_o}{k_B T}} \quad (6)$$

where f(J) accounts for the dependence of the vortex velocity on the current density. Now, let's suppose that for J=J* pinning is ineffective. We expect that, for J≥J*, the vortex velocity takes the form

$$\bar{v} = f(J) \quad (7).$$

For J=J*, there is a sudden increase in the dissipated power that, from Eqs.(5), (6) and (7), can be written as

$$\Delta P = I^* n_m L \phi_o f(J^*) \left(1 - e^{-\frac{U_o}{k_B T}}\right) \qquad (8)$$

where I* is the transport current at the transition. We have

$$\frac{\Delta P}{P^*} = e^{\frac{U_o}{k_B T}} - 1 \qquad (9)$$

where P* is the power dissipated just below J*. Below $T_{up}$, we have $\frac{U_o}{k_B T} \geq 4\pi$ [see Eq.(4)] and $\frac{U_o}{k_B T}$ keeps a large value up to a few Kelvin below $T_c$ (see Fig.3). Then, the ratio $\frac{\Delta P}{P^*}$ is large. Although in their work, thermal runaway is due to the self-heating of the sample without any intrinsic change in the vortex dynamics, Viña *et al.* in Ref.[19] have pointed out that a thermal runaway can be ignited only if $\frac{dP}{dt}$, the dissipated power rate, is large enough. In the range of temperature where $\frac{\Delta P}{P^*}$ is large, $\frac{dP}{dt}$ is large, in particular if the measurements are carried out with short current pulses as in the experiments of González *et al.* and we can expect thermal runaways. For T≈$T_c$, $U_o$ goes to zero as T goes to $T_c$ as seen in Fig.3 and we expect that $\frac{\Delta P}{P^*}$ is low. The increase in the dissipated power rate is not large enough to ignite a thermal runaway and no discontinuity in the CVCs is expected. We now determine the value of I*.

**II-C  Determination of the current at the transition to the normal state (T<$T_{up}$)**

From transport measurements, Goupil *et al.* [51] have inferred that depinning of an individual vortex line in YBCO samples occurs if the total force $\vec{F}$ acting on the vortex is equal to the maximum value of the pinning energy gradient. According to Brandt, [52,53] the pinning energy is an elastic energy.  Then, the maximum intensity of force $\vec{F}$ can be written as

$$F = J_{core} \nu \phi_o d = 2\frac{U_o}{\bar{\delta}} \qquad (10).$$

In Eq.(10),  $J_{core}$ is the current density carried by the vortex core and includes possible contributions due to the screening currents of other vortices. From Eqs.(3), (4) and (10), $I_{core}$, the current across the vortex core and $J_{core}$ can be written respectively as

$$I_{core} = 4 I_J \qquad (11a)$$

and

$$J_{core} = 4 J_c \qquad (11b)$$

since current $I_J$ flows across the weak links in the critical state. Now, let's examine the screening currents carried by the weak links. We assume that i) along the TB, the vortex screening current flowing across a weak link is equal to $\pm I_J$, as suggested in Section II-A and ii) the screening current is carried by weak links over distance $2\lambda$, where $\lambda$ is the film penetration depth. If a weak link carries the screening currents of two vortices, the resulting current is equal to $2I_J$ if the screening currents flow in the same direction. This is the case either if the weak link carries the screening currents of both a vortex and an antivortex (Fig.4a) or if the weak link carries the current of two vortices whose cores are less than $\lambda$ apart and is not located between the cores (Fig.4b). If the screening currents flow in opposite directions, the resulting current is equal to zero. This is the case either if the weak link is located between the cores of two vortices that are less than $\lambda$ apart or if the distance between the vortex cores is larger than $\lambda$, as shown in Fig.4c. Let's consider a sketch of the vortices lying in the vicinity of the annihilation line along a TB. In the critical state (Fig.5a), the TB is completely penetrated by the vortices and the antivortices. The mean distance between the vortex cores is equal to $2\lambda$ and there is no vortex interaction. Each weak link carries a screening current equal to $\pm I_J$, except the weak links at the vortex cores that carry no current and the weak link located along the annihilation line that carries a total screening current equal to $2I_J$, due to the vortex and the antivortex located at $\pm\lambda$. Above $J_c$, the vortices are in motion and, as confirmed by scanning SQUID microscopy observations, the vortex density increases with the transport current [42]. The distance between the cores of the two nearest neighbors of the annihilation line is not a constant. If this distance is less than $\lambda$, the core of each of these vortices carry current $I_J$ due to the screening current of the other one. For $J=2J_c$, the mean distance between the vortex cores is equal to $\lambda$. The total screening current flowing across the vortex cores is equal to zero, except across the cores of the vortices located at $\pm \lambda/2$ that, due to their next neighbors, carry a total screening current equal to $2I_J$ (Fig.5b). For $J=3J_c$, the mean distance between neighboring vortices is equal to $2\lambda/3$. The total screening current flowing across the vortex cores is equal to zero, except across the cores of the vortices located at $\pm\lambda/3$ that carry a screening current equal to $2I_J$ due to their next neighbors (see Fig.5c). However, due to its first and second neighboring vortices, the weak link located on the annihilation line carries a total screening current that is equal to $4I_J$, the depinning value. As the current goes on increasing, more weak links carry current $4I_J$ and we expect a sudden vortex depinning when current $4I_J$ flows across the cores of the vortices neighboring the TBs intersections in the vicinity of the annihilation line. Since the TBs intersections are only a few tens of nanometers apart [29], this occurs for a current density very near $3J_c$. It is reasonable to assume that this depinning process, possibly triggering a SOC chain reaction, ignites a thermal runaway as suggested in section II-B.

As a conclusion for this section, the strip enters the normal state if the current flowing across each weak link along the TBs is equal to $3I_J$ and current I* takes the form

$$I^* = 3I_J \overline{N} = 3I_J \frac{w}{\overline{\delta}} = 3I_c \qquad (12).$$

In the next section we show how J* for a given sample can be determined from the current-voltage curves measured in the vicinity of the critical state.

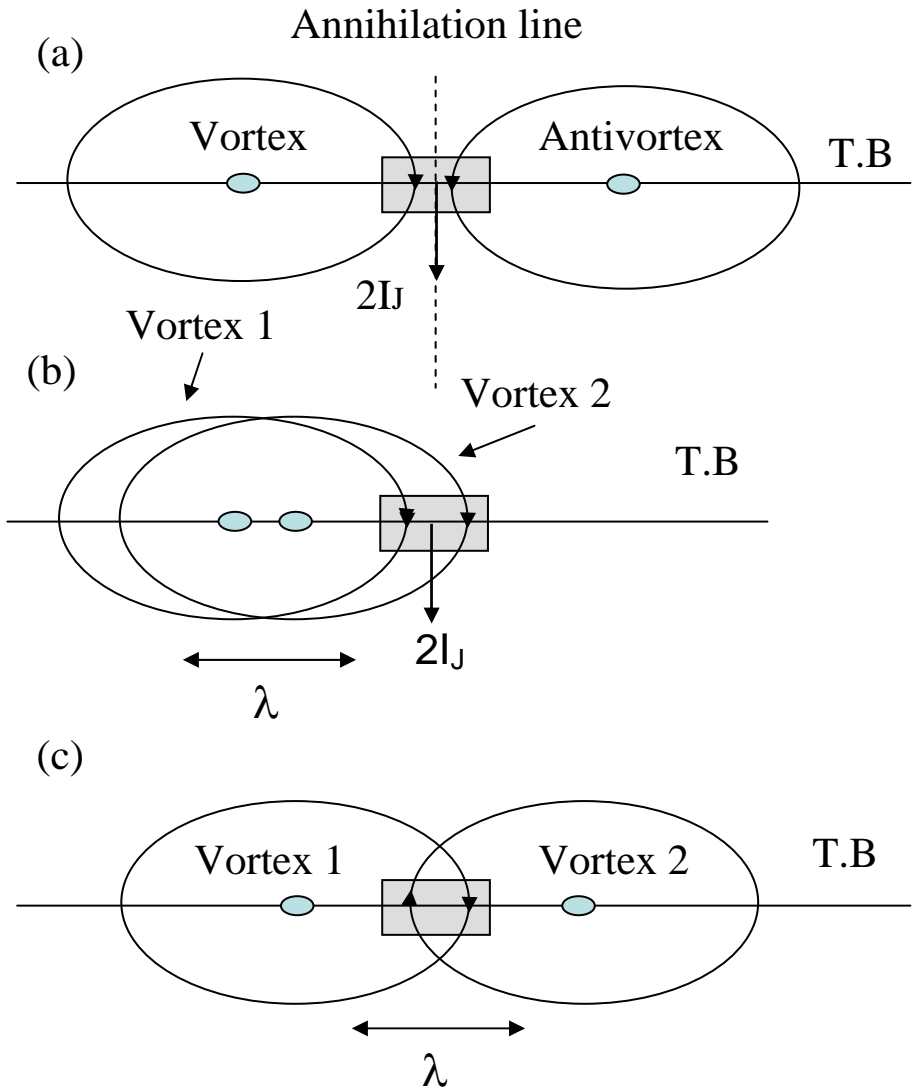

Fig.4 : Schematic representation of a weak link carrying the screening currents of two neighboring vortices. The screening current of each vortex is carried by the weak links over distance 2λ. The ellipses represent screening current lines and the grey areas in the ellipses represent the vortex cores. The rectangular grey area represent the weak link and the straight line the twin boundary, respectively; (a) the weak link carries the screening currents of both a vortex and an antivortex; (b) the weak link carries the screening currents of two vortices whose cores are less than λ apart and is not located between the cores; (c) the weak link carries the currents of two vortices whose cores are more than λ apart.

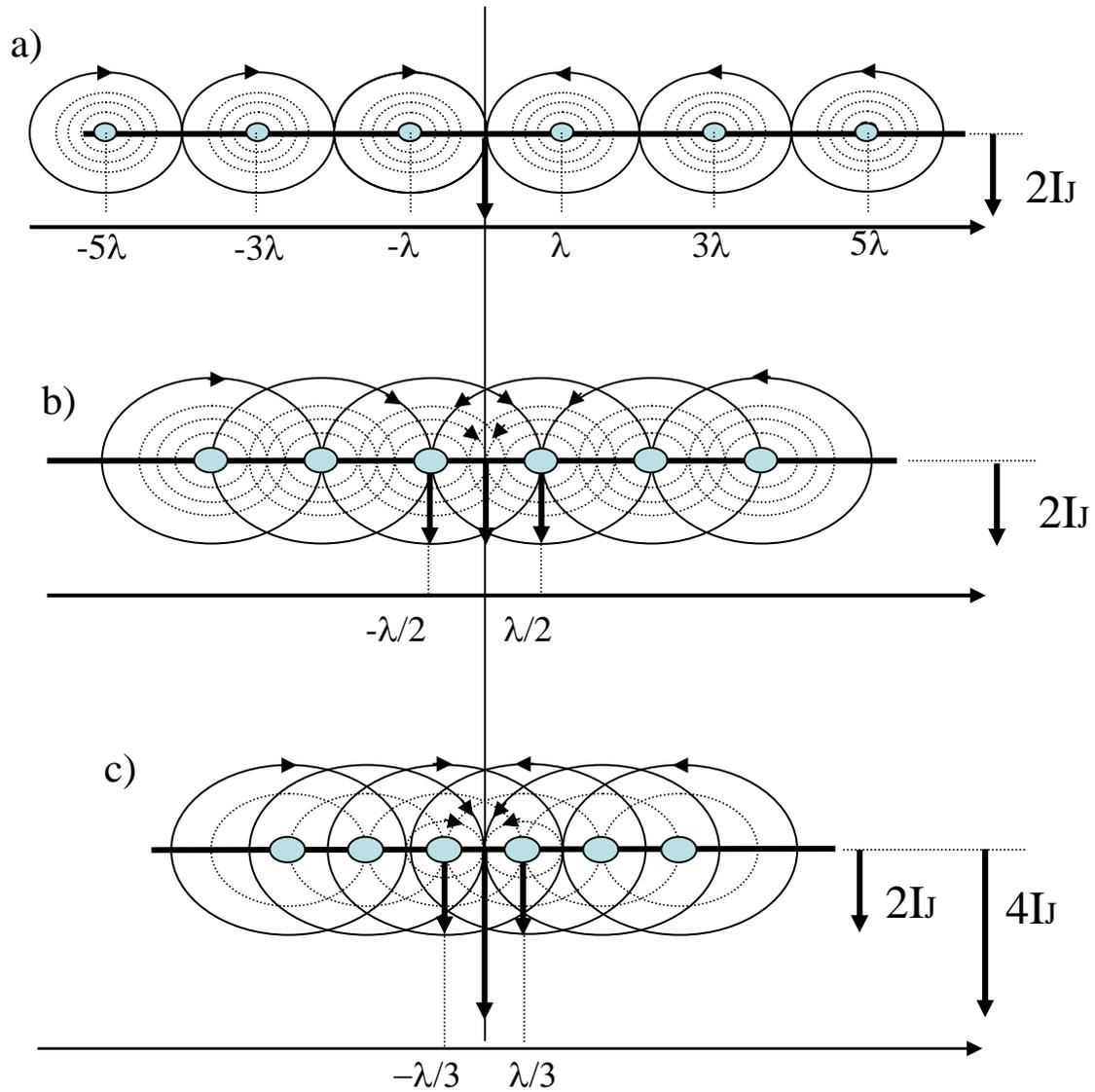

Fig.5 : Schematic representation of the vortices and antivortices lying along a twin boundary in the vicinity of the annihilation line for different current values : (a) I =$I_c$; (b) I=2$I_c$; (c) I=3$I_c$. The ellipses represent screening current lines and the grey areas in the ellipses represent the vortex cores. The thick and the vertical lines represent the twin boundary and the annihilation line, respectively. Only the screening currents flowing across the weak links located on the annihilation line and at the vortex cores are shown (full arrows).

**II – C  Determination of the value of J* from the current-voltage curves measured at low current.**

In Ref.[21], assuming that they are measured with a current low enough to neglect vortex interactions and no applied field, we have proposed a modified form of the Kim-Anderson expression to fit the CVCs. With electric field E and current density J, this expression takes the form

$$E = \rho_A (J - J_c) \exp\left(-\frac{U_o}{k_B T}\right) \sinh\left(\frac{J}{J_\delta}\right) \quad (13)$$

with

$$\frac{J}{J_\delta} = \frac{W}{k_B T} \quad (14).$$

In Eqs.(13) and (14) $\rho_A$ and $J_\delta$ are fitting parameters and $W = J v \phi_o d \overline{\delta}$ is the work carried out by the force due to the transport current acting on a vortex line that moves over distance $\overline{\delta}$. The difference between Eq.(13) and the Kim-Anderson expression [Eq.(1)] arises from the fact that Eq.(1) takes into account the vortices created by an external source and neglects the auto-induced vortices, while Eq.(13) considers auto-induced vortices only. From the expressions of W and Eqs. (3), (12) and (14) we can write relations between $J_c$, $J^*$ and $J_\delta$. We have respectively

$$J_c = 2\pi \nu J_\delta \quad (15)$$

and

$$J^* = 6\pi\nu J_\delta \qquad (16).$$

In addition, from Eqs.(4) and (15), $U_o$ can be written as a function of $J_c$ and $J_\delta$. We have

$$U_o = 2k_B T \frac{J_c}{J_\delta} \qquad (17).$$

The activation energy can also be computed from the values of $\rho_A$ and $J_\delta$. In principle, Eq.(17) is valid for $T<T_{up}$ only. However, the results published in Ref.[21] show that the behavior and the values of $U_o(T)$ obtained with Eq.(17) are similar to those obtained using $\rho_A$ and $J_\delta$ in the whole range of temperature. This is consistent with the considerations on the vortex regime above $T_{up}$ detailed in section IV, that suggest that Eq.(17) is also valid in this domain of temperature. We'll use Eq.(17) in what follows to determine $U_o$, because the CVCs of González *et al.* include few measurements at low current densities. This results in large errors in $\rho_A$ and in the determination of $U_o$ with expressions including $\rho_A$. We now compare the experimental results of González *et al.* to the results of section II.

**III – Comparison with experimental results**

González *et al.* measured the CVCs of two c-axis-oriented YBCO films called Sy116 and Sy3 whose thickness is equal to d=150nm and d=190nm, respectively. The films were grown on (100) SrTiO$_3$ substrates by high pressure DC sputtering. Details about the growth technique and the characterization of the films are described elsewhere [54]. Microbridges 10 μm wide were patterned by photolithography and wet chemical etching. The CVCs were measured at different temperatures in a four-probe configuration using either isolated pulses or stepped ramps with a step duration in the 1 ms range. The results obtained with the ramps did not significantly differ from those obtained using isolated pulses, as samples seemed to reach an almost steady state in less than 1 ms. In fact, the voltage was checked to be stable for square pulses around 30 ms long, up to a current density at least equal to 0.98J*, in agreement with previous works [54]. Only for currents very close to the discontinuity a progressive increase of the voltage signal, probably due to the heating of the sample, was observed.

In this section we first show that the CVCs can be fitted with Eq.(13) and we compare their shape near the transition to the predictions of section II-B. Then, we compare the values computed for the current flowing across each weak link in the critical state (Eq.2), $J_c$ [Eq.(15)] and J* [(Eq.(16)] to the experimental ones.

### III-A  Fitting of the current-voltage curves at low current

Fig.6 shows the experimental measurements carried out at low current on film Sy116 at 76.2K and the fitting curve obtained with Eq.(13). The inset shows the complete E(J) curve on a semi-logarithmic scale. This curve shows an almost linear part (on the semi-logarithmic scale) below an upper current density $J_2$ followed by a part with a different curvature between $J_2$ and J*. The experimental values could be fitted with Eq.(13) for J<$J_2$ only. This suggests that vortex interactions are different from that described in Fig.2 above this current range.

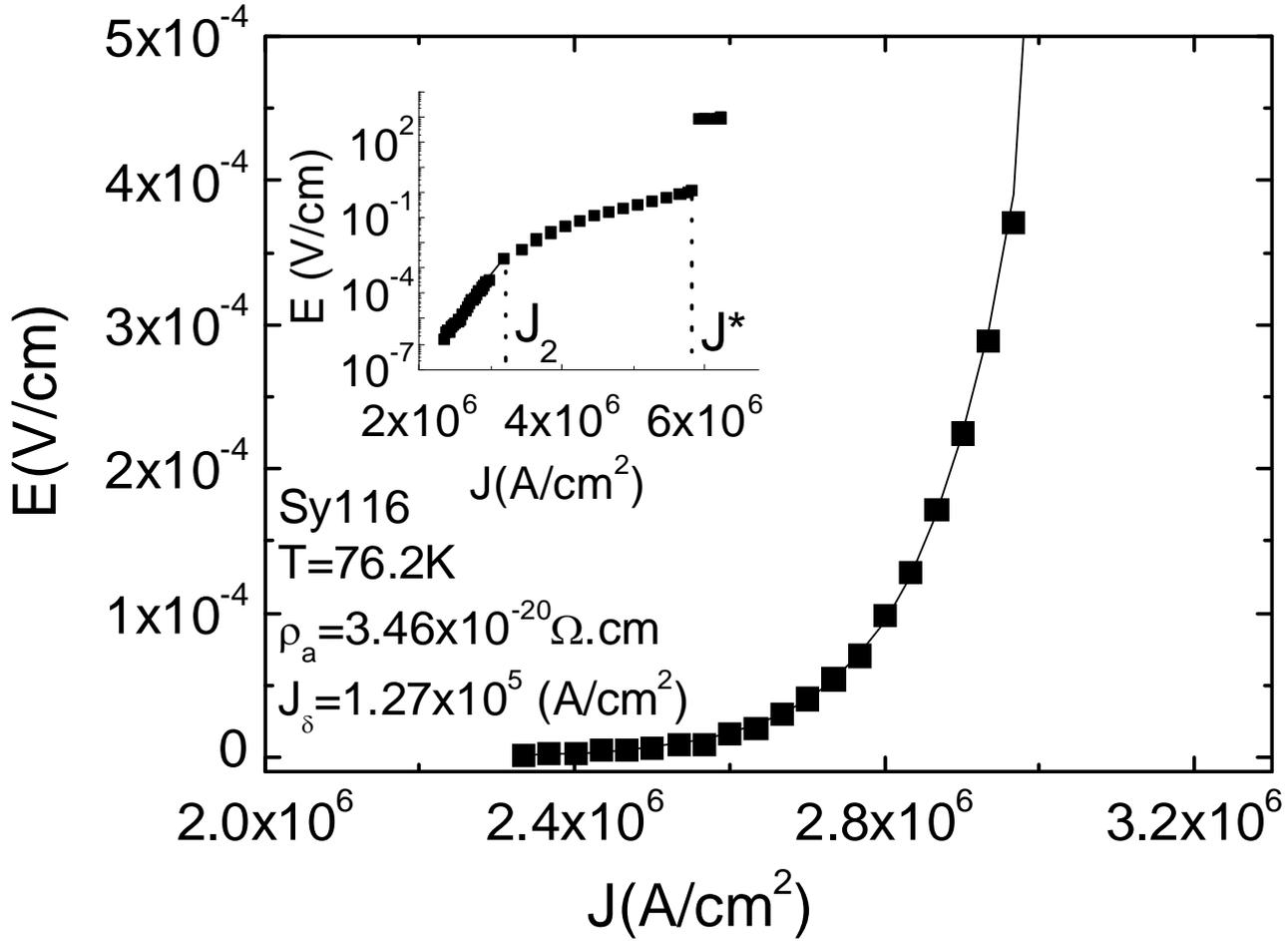

Fig.6 : Current-voltage measurements carried out on film Sy116 at 76.2K (full squares) and fitting curve at low current density obtained with Eq.(13) (solid line). J is the current density flowing in the sample and E the measured electrical field. The values of the fitting parameters are given in the figure. The inset shows the complete E(J) curve on a semi-logarithmic scale. The E(J) curve could be fitted with Eq.(13) below current density $J_2$ only (solid line). J* is the current density at the transition to the normal state.

**III-B  Current-voltage curves at the transition to the normal state**

Fig.7 shows the CVCs measured in the vicinity of $T_c$ on film Sy116. Film Sy3 shows similar characteristics. In the investigated range of temperature the CVCs show no voltage jumps, except at the transition to the normal state. This shows that this transition is not due to PSCs. The CVCs show a jump for $T \leq 88.9K$. For $T = 89.3K$ and $T = 89.7K$ the CVCs show a step but are continuous. The CVCs at 90K, 90.4K and 90.6K show a large linear part without any visible anomaly. Fig.8 shows the $U_o$ values that can be determined with Eq.(17). As previously pointed out, it was difficult to determine this quantity with a good accuracy because few measurements were carried out at low current densities. As expected the activation energy goes to zero as T increases above $T_{up} \approx 87.2K$ ($T_{up} \approx 88K$ for film Sy3). Fig.9 shows the $\frac{\Delta P}{P*}$ ratios computed from the $U_o$ values. Because of the lack of accuracy in the determination of $U_o$, only qualitative information can be obtained from this figure. In the lower temperature range, the order of magnitude of $\frac{\Delta P}{P*}$ is too large to be realistic. However, these high $\frac{\Delta P}{P*}$ values suggest that there is no regime of free vortex motion in the film because the large increase in the dissipated power at J=J* ignites a thermal runaway process immediately after depinning. In the vicinity of $T_c$, from the trend of the $\frac{\Delta P}{P*}$ curve at lower temperatures, we can infer that $\frac{\Delta P}{P*}$ is low. This can explain that the transition to the normal state is continuous. In the case of the measurements carried out at T ≥90K the CVCs shape suggests that the film is in the flux flow regime in a large domain of the current values. This means that the vortices move freely along the TBs and that there is a smooth transition from the thermally activated to the flux flow regimes. This behavior is consistent with the near zero values of $U_o$ and $\dfrac{\Delta P}{P*}$ expected from Figs.8 and 9 in this domain of temperature. No transition to the normal state is visible. This transition exists certainly, but for a current value outside the range of the measurements and, since the vortex velocity is very high in the flux flow regime, a LO process can't be excluded.

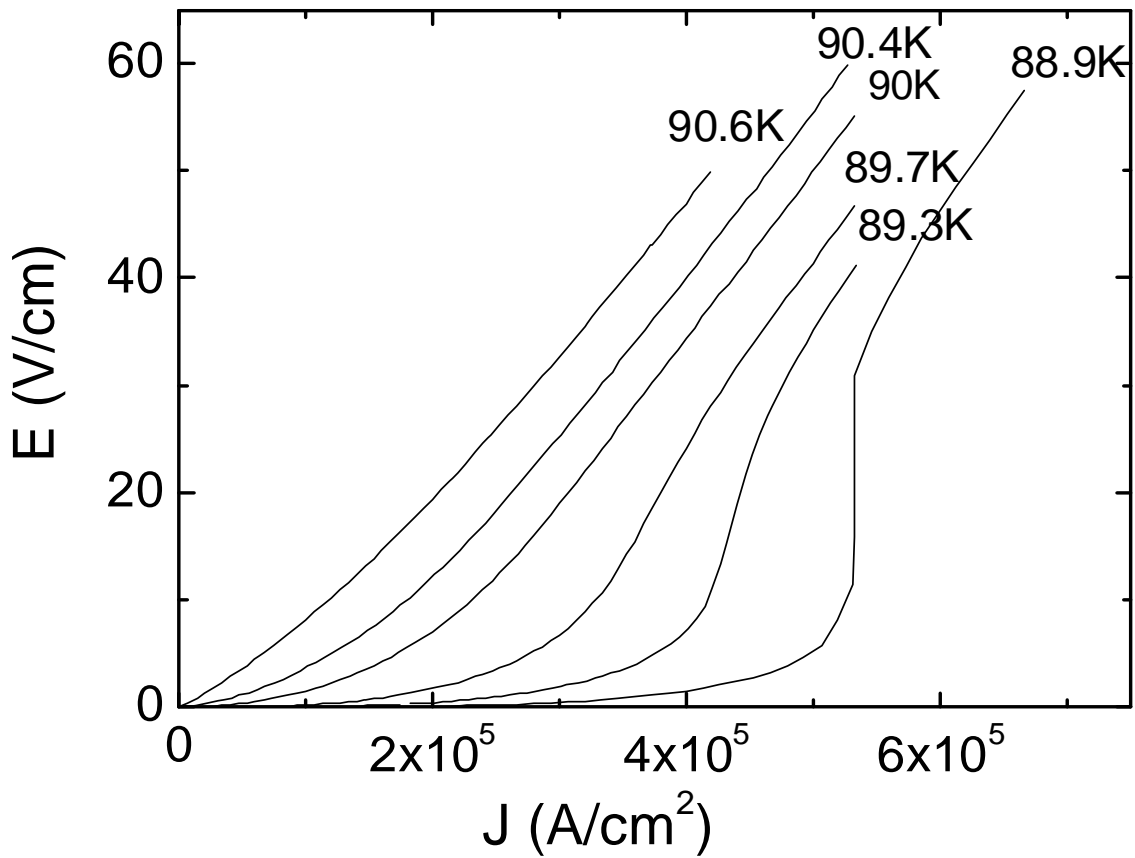

Fig.7 : Current-voltage curves measured on film Sy116 in the vicinity of $T_c$.

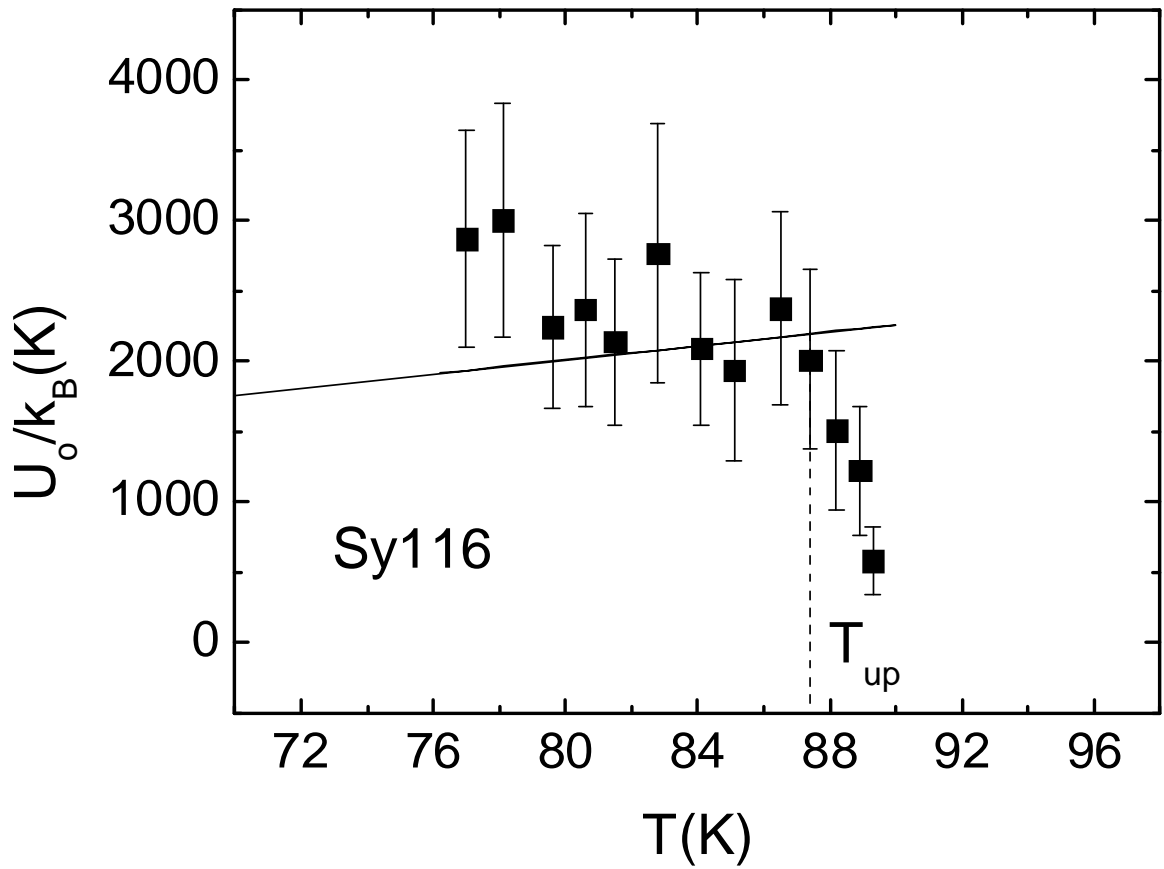

Fig.8 : Activation energy of film Sy116 as a function of the temperature. The activation energy was determined according to Eq.(17). The straight line represent the values computed with Eq.(4) and ν=2.

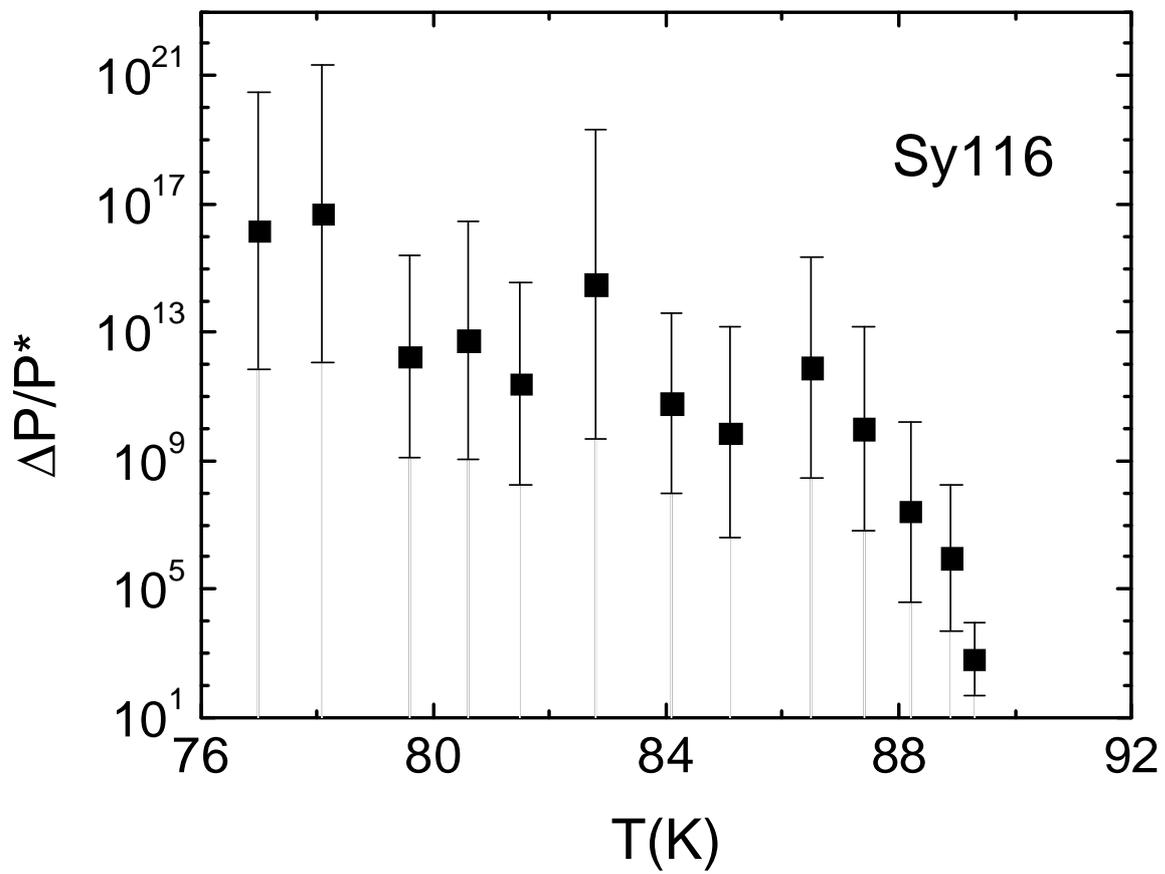

Fig.9 : Relative increase in the dissipated power of film Sy116 at J=J* as computed from the activation energies in Fig.8.

**III-D – Weak link current, critical current density and current density at the transition to the normal state**

### 1 – Weak link current

Gonzàlez *et al.* have determined with accuracy the $J_c$ values of films Sy116 and Sy3 from CPL fittings of the E(J) curves. The current flowing across each weak link in the critical state can be calculated as $I_{wl} = J_c d \bar{\delta}$. The $I_{wl}$ values obtained for both samples are compared to the values calculated with Eq.(2) in Fig.10, while the inset shows the corresponding $\bar{\delta}$ values. The $\bar{\delta}$ values increase as the temperature increases, as expected. The experimental $I_{wl}$ values tend to zero as T goes to $T_c$. In spite of the scattering due to the error in the determination of $\bar{\delta}$, the $I_{wl}$ values are in the range of $\frac{2\pi k_B T}{\phi_o}$ below the $T_{up}$ temperature of each film, as predicted by the model.

### 2 – Critical current density

In order to compare the experimental $J_c$ values to those computed with Eq.(15), the $\nu$ value of each film must be determined. The $U_o$ values computed for film Sy116 (see Fig. 8) suggest that $\nu=2$ for this film. Similarly, they suggest that $\nu=1$ for film Sy3. The $J_c$ values determined by Gonzàlez *et al.* are compared in Fig.11 with the values obtained with Eq.(15) taking these values for $\nu$. The agreement between the values determined by Gonzàlez *et al.* and those computed with Eq.(15) is good for both films below $T_{up}$.

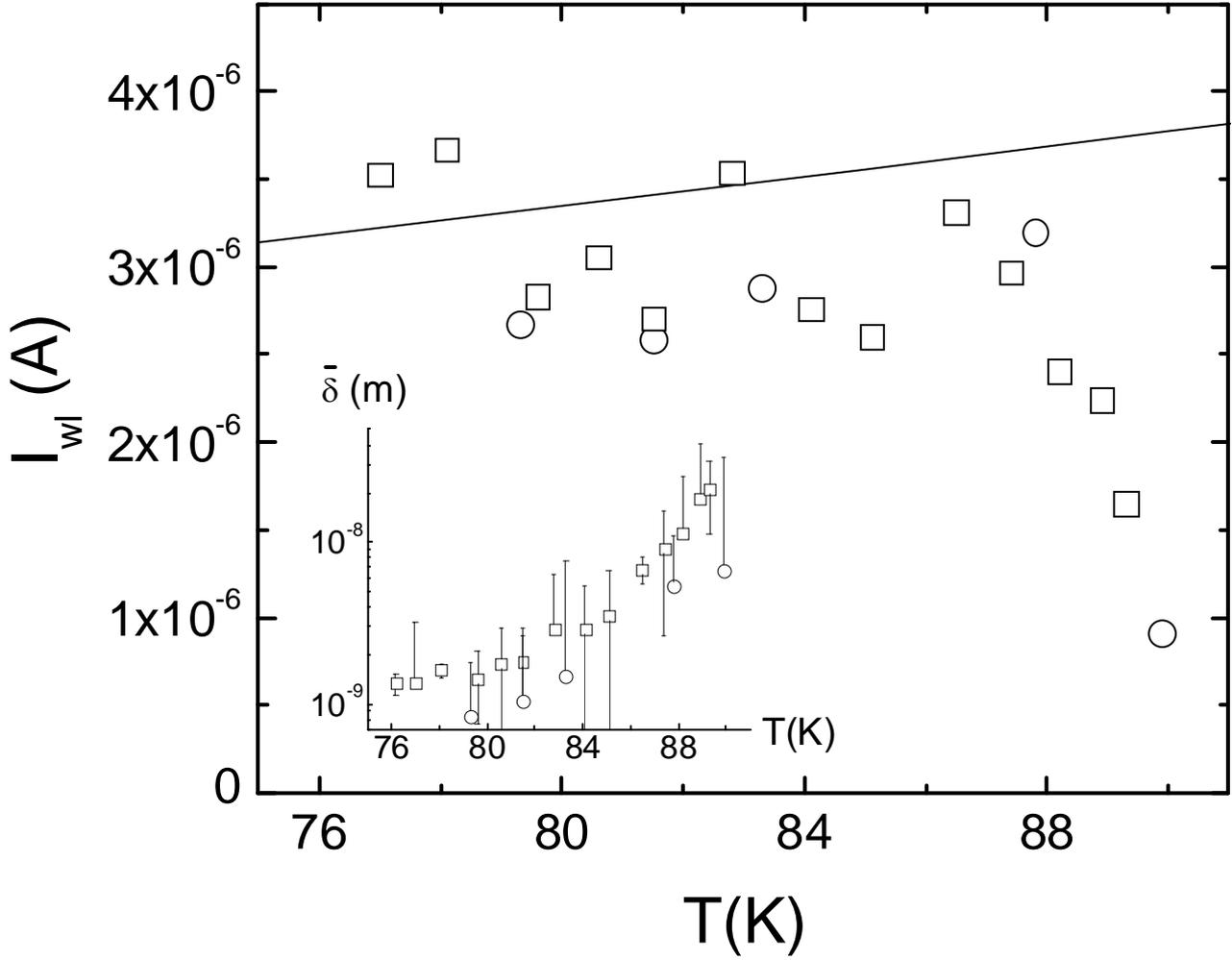

Fig.10 : Current $I_{wl}$ carried by each weak link of films Sy3 (circles) and Sy116 (squares) in the critical state as a function of the temperature, according to the expression $I_{wl} = J_c d \bar{\delta}$. In this expression $J_c$ is the critical current density, $\bar{\delta}$ the mean length of the weak links along the TBs and d the film thickness. The solid line shows the predicted values $I_{wl} = \dfrac{2\pi k_B T}{\phi_o}$ (Eq.2). The inset shows the $\bar{\delta}$ values.

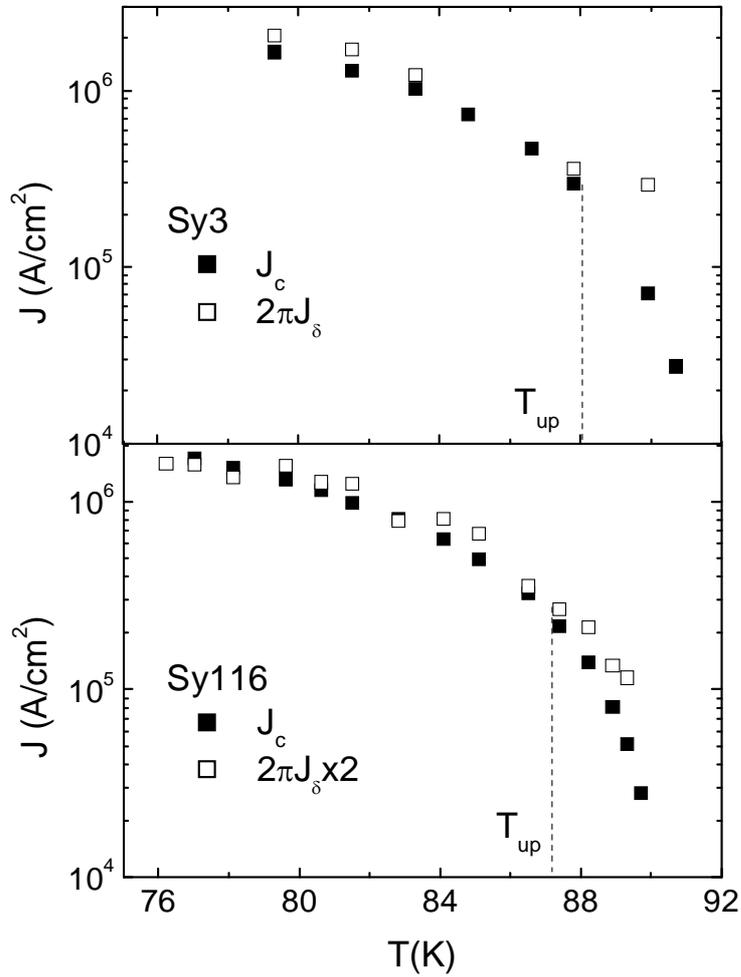

Fig.11 : Critical current densities of films Sy3 and Sy116 as determined by González *et al.*[18] (full squares) and as computed with Eq.(15) and $\nu=1$ for film Sy3 and $\nu=2$ for film Sy116 (open squares).

**3 – Current density at the transition to the normal state**

Current density J* was determined by Gonzàlez *et al.* from a deviation criterion of the experimentally measured E values from the background dissipation given by different models. They noted only a weak sensitivity to the model choice. The J* values obtained using the CPL are compared in Fig.12 to the values computed with Eq.(16) taking the same ν values as for $J_c(T)$. The quantity $3J_c$ for both films is also reported. In the case of film Sy3, there is a good agreement between J*, $6\pi\nu J_\delta$ (ν =1) and $3J_c$ in the whole range of the J* measurements. However, no J* value was available above $T_{up}$, since this film does not show thermal runaway in this range of temperature. In the case of film Sy116 there is a good agreement between J*, $6\pi\nu J_\delta$ (ν=2) and $3J_c$ below $T_{up}$. Above $T_{up}$, the agreement between J* and $6\pi\nu J_\delta$ is still very good, but J* is different from $3J_c$.

The results obtained in this section show that the dependence on temperature of the shape of the CVCs at the transition to the normal state is consistent with that of the vortex pinning energy. They also confirm that in the critical state the weak link energy is equal to $k_B T$ as suggested by the model. In addition, $J_c$ can be reproduced with Eq.(15) below $T_{up}$, as expected. Surprisingly, as shown in Fig.12, the J* values of film Sy116 can be reproduced with Eq.(16) in the whole range of temperature although we have $J^*/J_c \gg 3$ at the highest measurement temperatures, as seen in Fig.13. This last point is discussed in section IV.

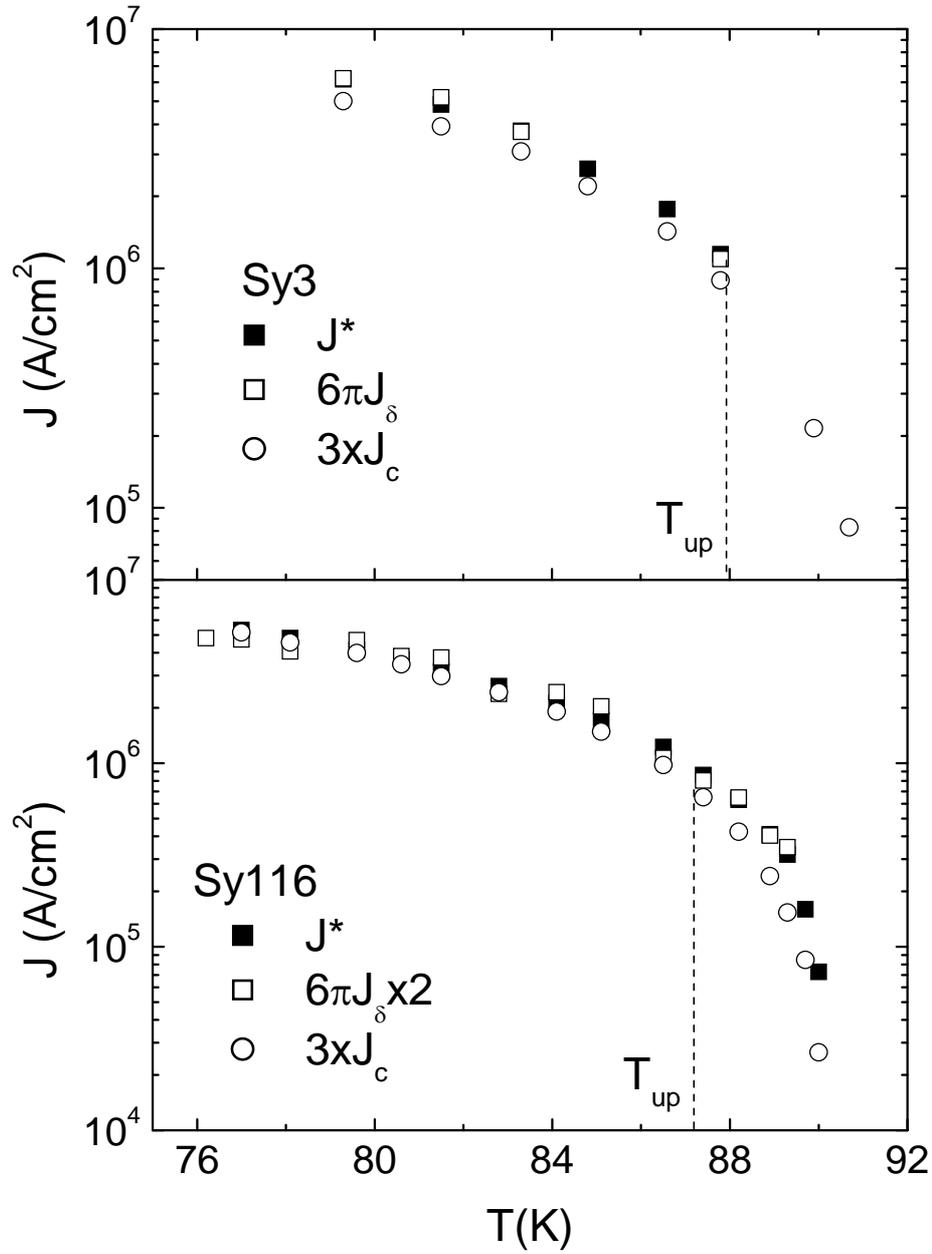

Fig.12 : Quantity $3 \times J_c$ (open circles) and current densities at the transition to the normal state as determined by González *et al.*[18] (full squares) and as computed with Eq.(16) and $\nu=1$ for film Sy3 and $\nu=2$ for film Sy116 (open squares).

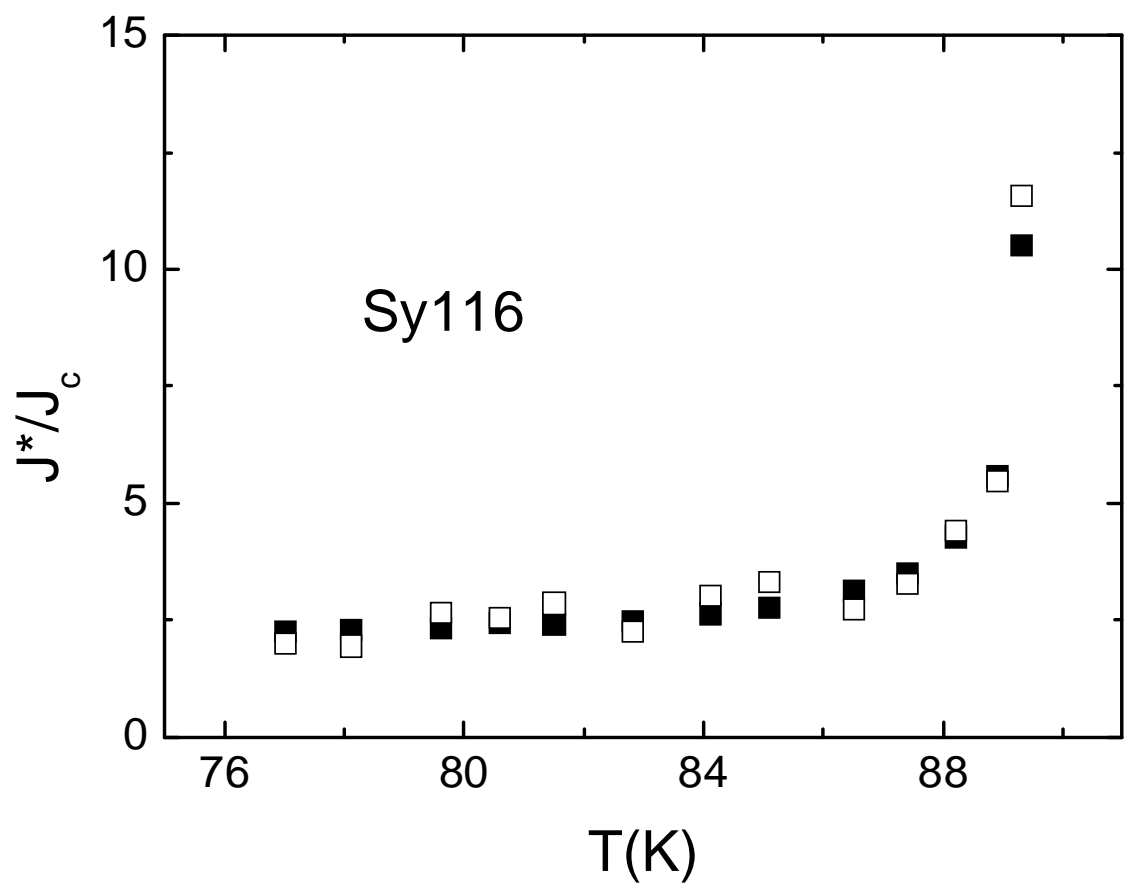

Fig.13 : Ratio $J^*/J_c$ for film Sy116. Full squares : experimental values ; open squares : ratios computed with the $J^*$ values given by Eq.(16).

**IV - Discussion**

In this section we discuss the relevance of the model as compared to other descriptions and we compare the results of the previous section to that obtained by other authors. We infer an extension of the model above $T_{up}$ from the experimental results.

**IV–A  Relevance of the model and comparison with other results**

As detailed in Sect.I, the transition to the normal state of YBCO films has been ascribed to different physical phenomena and different models have been proposed. Gonzàlez *et al.* have stressed out that the models based on vortex dynamics that assume power laws for the CVCs have too many free parameters to be considered as reproducing unequivocally the experimental results. In Sect.III-C, we have pointed out that in films Sy3 and Sy116 the transition could not be due to PSCs. Phase shifts centers have been observed in YBCO films or in films of the same family deposited on MgO substrates. However, in addition to twin boundaries these films include domains with 45° misorientations [55] that play probably a role in their response to the applied current. There are certainly defects in the films of Gonzàlez *et al.* that could behave as heating points. However, taking into account the thermal impedance of the substrate and the temperature dependence of the electrical resistivity of the films only, a very good agreement between experimental and computed J* values can be obtained [19, 20] . This shows that it is not necessary to suppose the existence of these defects to ascribe the transition to thermal instability. However, thermal heating can't account for the non-discontinuity of the transition in the near vicinity of $T_c$. More experimental work, especially measurements with very short current pulses to limit sample heating, will be necessary to separate the effects of thermal heating from those related to vortex dynamics.

In section II-C we have proposed expressions showing that $J_c$ and J* are closely related. This conclusion was previously suggested by Xiao *et al* [9], that have pointed out that the temperature and magnetic field dependence of J* and $J_c$ show similar scaling relations. The experimental measurements carried out on films Sy116 and Sy3 have confirmed that J* is equal to $3J_c$ up to a few Kelvin below $T_c$. This result was verified by other studies. Measurements carried out at 86.9K by Peterson *et al.*[17] on a microbridge showed that the ratio of the current required to maintain this sample in the normal state to the critical current is equal to 3 if no magnetic field is applied. Decroux *et al.* [1] have observed that a very fast transition into a highly dissipative state occurs in YBCO films at 77K for a current density larger than $3J_c$.

**IV-B – Vortex dynamics above $T_{up}$**

For $T>T_{up}$, $U_o$ is a decreasing function of the temperature and we have $J_c < 2\pi\nu J_\delta$ as observed in Fig.11. Since the density of superconducting pairs decreases as the temperature increases, we can suppose that, in this range of temperature, the TBs include weak links whose maximum superconducting current is smaller than $I_J = \dfrac{2\pi k_B T}{\phi_o}$. These "dead" weak links carry no current in the superconductive and the critical states of the film and we have $I_c < I_J \overline{N}$, although in the critical state the TBs are completely penetrated by the vortices, as represented in Fig.5a. Since there are weak links randomly located along the TBs that carry no screening current, it is probable that the weak links located at the vortex centers are not the only weak links included in the vortex cores. We expect that the cores include also the dead weak links located near the vortex centers (see Fig.14). If $\overline{r}_c$ is the mean length of the vortex cores along the TBs, the ratio $\dfrac{\overline{\delta}}{\overline{r}_c}$ is equal to unity for $T \leq T_{up}$. This ratio goes to zero as T goes to $T_c$ since, as the temperature increases, the number of dead weak links and the size of the cores increase. As an approximation, we assume that $\frac{\overline{\delta}}{r_c}$ is equal to the mean ratio of the superconducting to the total number of weak links in the TBs. Then, the critical current can be written as

$$I_c = \frac{\overline{\delta}}{r_c} \overline{N} I_J \qquad (18).$$

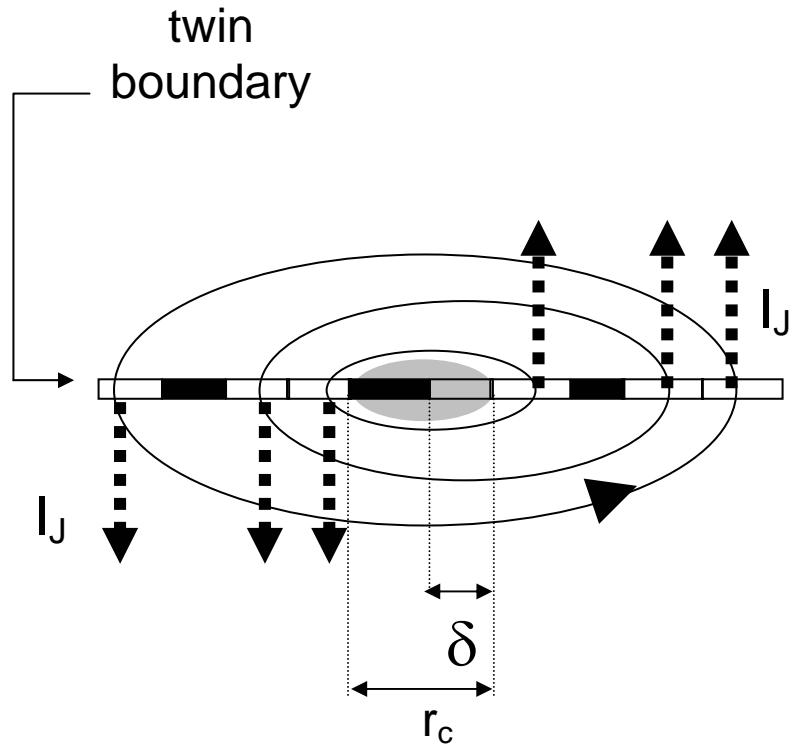

Fig.14 : Schematic representation of a vortex located along a twin boundary in the critical state at temperature $T>T_{up}$. The rectangular white and black areas represent weak links whose maximum superconducting current is respectively larger and smaller than $I_J$. The ellipses represent screening current lines and the oval grey area the vortex core. The quantities $\delta$ and $r_c$ are the weak link length and the length of the vortex core along the TB, respectively. Only those weak links whose maximum superconducting current is at least equal to $I_J$ carry screening currents. The vortex core is not restricted to a single weak link since it can include weak links with a maximum supercurrent smaller than $I_J$. If current $I_J$ enters the core it sweeps flux $\dfrac{\delta}{r_c}\nu\phi_o$ instead of flux $\nu\phi_o$ below $T_{up}$

Let's consider two vortices located near a TB intersection. As for $T<T_{up}$, the displacement of one of the vortices over distance $\delta$ is possible only if screening current $I_J$ enters both its core and the core of the other vortex (see Fig.2). However, since the mean flux threading each weak link included in the vortex core is equal to $\frac{\overline{\delta}}{r_c}\phi_o$, the energy barrier is

$$U_o = 2I_J \frac{\overline{\delta}}{r_c} \nu \phi_o \qquad (19).$$

The critical current $I_c$ and the activation energy $U_o$ given by the above equations go to zero as T goes to $T_c$, as observed experimentally. It is easily seen from the expression of W and Eqs.(14), (18) and (19) that $U_o$ can also be written in the form of Eq.(17). For $J>J_c$, there is a voltage drop between the two banks of the TBs and hence, the transport current can be supposed to flow across all the weak links, while the screening currents flow only across those weak links that can carry a pair current at least equal to $I_J$. Vortex depinning occurs if Eq.(10) is satisfied. We have

$$F = J_{core}\nu\phi_o d = \frac{I_{core}}{\bar{r}_c}\nu\phi_o = 2\frac{U_o}{\bar{\delta}} = 4\frac{I_J}{\bar{r}_c}\nu\phi_o \qquad (20)$$

and

$$I_{core} = 4I_J \qquad (21).$$

Eq.(21) sets a condition identical to that found for $T<T_{up}$. If all the weak links carry transport current $3I_J$, the situation is identical to that described in Fig.5c. However, for $T>T_{up}$, the current density at the transition is larger than $3J_c$ since not all the weak links carry current $I_J$ in the critical state.

## V - Summary and conclusions

We have proposed a description accounting for the results obtained by Gonzàlez *et al.*[18]. This description is valid for YBCO films where vortex dynamics is at first related to the existence of twin boundaries, as this is the case for c-axis oriented films deposited on $SrTiO_3$ substrates. According to this model, the vortices are in motion along the TBs that, due to disorder, behave as rows of weak links. Vortex pinning occurs at the TBs intersections. At low temperature (below $T_{up}$) all the weak links carry the same current $I_J(T)$ in the critical state. The transition to the normal state occurs for $J^*=3J_c$. It is triggered by vortex depinning at the TBs intersections neighboring the annihilation line. Vortex depinning is responsible for the discontinuities observed in the CVCs because it causes a strong increase in the dissipated power rate. Above $T_{up}$, some weak links carry no current in the non resistive regimes while all the weak links carry current for $J>J_c$. This results in a $J^*$ value larger than $3J_c$. In the near vicinity of $T_c$, the vortex pinning energy is almost zero and vortex depinning does not induce an increase in the power rate large enough to cause a thermal runaway. In this domain of temperature, the vortices can be driven in the flux flow regime. The results published by González *et al.* are in a good agreement with the predictions of this description that is not in contradiction to the suggestions that LO and SOC processes are involved in the transition to the normal state. However, the LO process is possible in the near vicinity of the critical temperature only, as suggested in Ref.[16], while the SOC process would occur at lower temperatures.

**Aknowledgments**

The authors are grateful to Félix Vidal and Jesús Maza for their suggestions and comments. They thank Maryvonne Hervieu for the help she brought them. T. G. and M. R. wishes to acknowledge support from the CICYT, Spain, under grants MAT2001-3272 and MAT2004-04364.

**Annex A     Weak link coupling energy in the critical state**

The TBs are highly disordered. It is reasonable to assume that, at a given temperature, in each weak link, regions exist where the current density is equal to the minimum value compatible with the existence of a tunneling pair current. Since the current density is uniform in short weak links, it is equal to this minimum value. The resulting current is the minimum current maintaining phase coherence across the weak link against thermal fluctuations. The corresponding weak link energy is equal to $k_BT$. Another argument for this proposition is that to lower their repulsive energy the vortices tend to occupy a surface as large as possible. However, for $I \leq I_c$ they are present in the region where the current flows only. As a result, the largest is this region, the lowest is the vortex repulsive energy. This condition is fulfilled if, in the region penetrated by the current, the weak links carry the lowest current enabling them to maintain the phase coherence across the TBs, i.e. if their energy is equal to $k_B T$. In addition, the Hamiltonian of the twin boundaries is also minimum in this case if we suppose that it takes the form $H = \sum_i E_{J,i}(1 - \cos\phi)$. In this expression $\phi$ is the difference in the phase of the order parameter between the superconducting banks and $E_{J,i}$ is the coupling energy of the i-th weak link in a twin boundary.